\documentstyle[epsfig,longtable]{aipproc}
\def\D0{D\O}
\def\ifmath#1{\relax\ifmmode #1\else $#1$\fi}%
\def\GeV{\ifmmode {\mathrm{ Ge\kern -0.1em V}}\else
                   \textrm{Ge\kern -0.1em V}\fi}
\def\MeV{\ifmmode {\mathrm{ Me\kern -0.1em V}}\else
                   \textrm{Me\kern -0.1em V}\fi}
\def\GeV{\ifmmode {\mathrm{ Ge\kern -0.1em V}}\else
                   \textrm{Ge\kern -0.1em V}\fi}%
\def\MeV{\ifmmode {\mathrm{ Me\kern -0.1em V}}\else
                   \textrm{Me\kern -0.1em V}\fi}%
\def\keV{\ifmmode {\mathrm{ ke\kern -0.1em V}}\else
                   \textrm{ke\kern -0.1em V}\fi}%
\def\eV{\ifmmode  {\mathrm{ e\kern -0.1em V}}\else
                   \textrm{e\kern -0.1em V}\fi}%
\def\GeVcc{\ifmmode {\mathrm{ \GeV/c^2}}\else
                     \textrm{Ge\kern -0.1em V/c$^2$}\fi}%
\def\MeVcc{\ifmmode {\mathrm{ \MeV/c^2}}\else
                     \textrm{Me\kern -0.1em V/c$^2$}\fi}
\newcommand{\SM}        {\mbox{SM}}
\newcommand{\GZ}        {\Gamma_{\mathrm{Z}}}
\newcommand{\Gll}       {\Gamma_{\ell\ell}}
\newcommand{\Ghad}      {\Gamma_{\mathrm{had}}}
\newcommand{\Gbb}       {\ifmath{\Gamma_{\mathrm{b\bar{b}}}}}
\newcommand{\Gnu}       {\Gamma_{\nu\nu}}
\newcommand{\Rl}        {R_{\ell}}
\newcommand{\shad}      {\sigma_{\mathrm{h}}^{0}}
\newcommand{\ALR}       {\mbox{$A_{\rm {LR}}$}}
\newcommand{\AFB}       {A_{\mathrm{FB}}}
\newcommand{\cAe}       {\mbox{$\cal A_{\rm e}$}}

\newcommand{\cAf}       {\mbox{$\cal A_{\rm f}$}}

\newcommand{\swsqeffl}  {\sin^2\!\theta_{\rm{eff}}^{\rm {lept}}}
\newcommand{\Rb}        {\ifmath{R_{\mathrm{b}}}}
\newcommand{\ee}        {\mbox{$e^+e^-$}}

\newcommand{\pbarp}     {\mbox{$\overline{p}p$ }}
\newcommand{\qqbar}     {\mbox{$\overline{q}q$}}

\newcommand{\lsim}{\mathrel{\raisebox{-.6ex}{$\stackrel{\textstyle<}{\sim}$}}}
\newcommand{\gsim}{\mathrel{\raisebox{-.6ex}{$\stackrel{\textstyle>}{\sim}$}}}

\begin{document}

\rightline{\vbox{\halign{&#\hfil\cr
&{\bf Fermilab-Conf-98/030 } \cr
&{\bf MADPH-98-1037} \cr
&{\bf hep-ph/9801407} \cr
&January 1998\cr}}}
\rightline{\qquad}

\title{Higgs Boson and $Z$ Physics \\ at the First Muon Collider}

\author{Marcel Demarteau$^*$ and Tao Han$^{\dagger}$}
\address{$^*$Fermi National Accelerator Laboratory, 
P.O.Box 500, Batavia, IL 60510\\
$^{\dagger}$Department of Physics, University of California, Davis, CA 95616\\
and\\
Department of Physics, University of Wisconsin, Madison, WI 53706}

\maketitle

\begin{abstract}
The potential for the Higgs boson and $Z$-pole physics 
at the first muon collider is summarized, based on the discussions 
at the \lq\lq Workshop on the Physics at the First Muon 
Collider and at the Front End of a Muon Collider\rq\rq .
\end{abstract}

\section*{Introduction}

Muon colliders offer a wide range of opportunities for exploring 
the physics within and beyond the Standard Model (SM). Because the muon 
mass is about 200 times larger than the electron mass, 
$s$-channel production of the Higgs boson, and its associated advantages 
with regard to measurements of Higgs boson properties, 
is one of the unique features of a muon collider. 
Since the muons are produced in a decay channel by moving pions, the 
muons naturally carry a longitudinal polarization of about 20\%. 
A collider allowing for adroit manipulation of the polarization and 
center of mass energy, combined with the prospect of luminosities in the range 
of $10^{32} < {\cal L} < 10^{33}$cm$^{-2}$s$^{-1}$ would make for a very 
powerful probe of the structure of the fundamental forces with 
unparalleled potential. 
In this report a summary of the Higgs and $Z$-pole physics, 
as well as some other aspects of the electroweak boson physics,
as discussed at the workshop on the physics at the 
First Muon Collider (FMC), is presented~\cite{part}.

\section*{Experimental Considerations}

The experiments at the $e^+e^-$-colliders LEP and SLC have shown that 
the calibration of the luminosity, beam energy and beam polarization is 
crucial for the physics results obtained. 
At LEP the luminosity is measured with small angle silicon based 
calorimeters, counting Bhabha events to a precision of 
${\delta{\cal L} \over {\cal L}} = 10^{-3}$. 
The Bhabha cross section has been measured down to angles of about 30~mrad
with respect to the beam direction. 
At the FMC, however, it is not clear if the muon Bhabha cross section can 
be measured down to small angles. The muons, with a lifetime of about 
2~$\mu$s in their rest frame, can circulate in the machine for only about 
800 turns at a center of mass energy ${\sqrt s} = 0.5$~TeV~\cite{fmc}. 
The electrons from the decay of the muon for a major problem. 
For $2 \cdot 10^{12}$ muons per bunch there are 
$3 \cdot 10^{5} / E_{\rm beam}$ decays per meter, 
with $E_{\rm beam}$ in TeV. Because the final focus is tuned to the beam 
energy of the muon beams, the decay electrons will be sprayed out over the 
interaction region. Current detector designs~\cite{lebrun} 
include an uninstrumented cone of $10^\circ$ - $20^\circ$ with respect to 
the beamline 
because of the large direct and induced backgrounds. 
It is thus unclear if a similar 
precision on the luminosity measurement can be obtained using muon 
Bhabha scattering or using an alternative method. For the discussions 
to follow it is assumed that at the FMC a precision on the 
luminosity measurement of ${\delta{\cal L}/ {\cal L}} = 10^{-3}$ 
is achievable.

The beam energy at LEP is 
measured most accurately using the technique of resonant 
depolarization which has an ultimate accuracy of about 200~keV. 
This calibration, however, cannot be performed very often since 
it takes a long time for the polarization to build up in the beam. 
Moreover, it cannot be done during a physics run and has been performed 
with separated beams only. These and other limitations, resulted in a
final uncertainty on $\sqrt{s}$ at LEP of about 1~MeV. 

At the FMC the natural polarization of the muons~\cite{fmc} 
provides a mechanism 
to measure not only the beam energy but also the polarization itself. 
The precession of the polarization vector with respect to the momentum vector 
is governed by the muon spin tune, $\frac{g-2}{2} \, \gamma$, 
which corresponds to 
the number of precessions in one turn around the ring. 
Coincidentally, for muons at the $Z$-mass the spin tune is almost exactly 
${1/ 2}$. Thus, the spin flips each turn at $\sqrt{s} \approx M_Z$.  
The energy spectrum of the decay electrons depends on the 
muon polarization.
By measuring the  average energy of the decay electrons each turn at a 
fixed point along the circumference of the machine a measure of the 
beam energy and the polarization can be obtained. The measured energy 
spectrum will exhibit an oscillatory behavior as function of turn number. 
The frequency and amplitude of the oscillations measure the beam 
energy and the polarization, respectively. 
Initial studies for a perfect planar machine geometry 
show that an absolute energy calibration at the statistical level of 
${\delta E / E } = 10^{-5}$   
is easily feasible~\cite{Raja}. 
Systematic effects will be the dominant sources of uncertainty and 
are being studied. 
A clear advantage of this procedure is that the measurements are done 
concurrently with physics runs, directly sampling the interacting muon 
bunches. 

The beam spread is controlled by the beam optics and a narrow band beam 
option at lower center of mass energies would provide a beam spread of 
$R = {\sigma(p)/ p} = 3 \cdot 10^{-5}$. 

\section*{Higgs Bosons}

The $\SM$ as it stands is incomplete. Many fundamental questions 
of nature are left unanswered. Among them the question of electroweak 
symmetry breaking takes a prominent position. In the $\SM$, 
the electroweak symmetry is broken spontaneously
through a fundamental Higgs doublet field, giving
rise to a single physical neutral Higgs boson ($h^0$).
In the minimal supersymmetric standard model (MSSM), there are
two neutral ${\cal CP}$-even Higgs bosons ($h^0,H^0$), two charged Higgs 
bosons ($H^\pm$), and a ${\cal CP}$-odd neutral Higgs boson ($A^0$). 
To understand the electroweak symmetry breaking and to explore
new physics beyond the SM, the study of the Higgs sector is of 
highest priority for future collider experiments \cite{Haber}. 

Although the Higgs boson masses are largely free parameters,
theoretical arguments indicate that there exists an upper
limit on the lightest Higgs boson mass, namely, 
$m_h<150$~\GeVcc~\cite{Haber}. 
The FMC then is unique in studying light Higgs bosons since
they can be produced through $s$-channel production at resonance, 
due to the sizeable coupling of a Higgs boson 
to muons \cite{Barger,Gunion,BBGH}. 

The resonance cross section for Higgs boson production is given by
\begin{equation}
\sigma_h\left(\sqrt{\hat s}\right) = { 4\pi\Gamma(h\to\mu\mu)\, \Gamma(h\to X)
\over \left(\hat s - m_h^2\right)^2 + m_h^2 \Gamma_h^2
}\,,
\end{equation}
where $\hat s$ is the c.m.\ energy squared, $\Gamma_h$ is the total
width, and $X$ denotes the final state. The Higgs coupling to fermions is
proportional to the fermion mass so the corresponding $s$-channel process is
highly suppressed at $e^+e^-$ colliders with respect to muon colliders. 
The cross section must be convoluted with the machine energy spectrum, 
approximated by a Gaussian distribution of width $\sigma^{}_{\sqrt s}$,
\begin{eqnarray}
\bar\sigma_h\left(\sqrt s\right) = \int \sigma_h\left( \sqrt{\hat s} \right)
 {\exp  \left[- \displaystyle  { \left( \sqrt{\hat s} - \sqrt s \right)^2 \over
\left(2\sigma_{\sqrt s}^2 \right)} \right] \over\sqrt{2\pi}
\,\sigma^{}_{\sqrt s} } \  d\sqrt{\hat s} \, .
\end{eqnarray}
The root mean square spread $\sigma^{}_{\sqrt s}$ in c.m.\ 
energy is given in terms of the beam resolution $R$ by
\begin{equation}
\sigma_{\sqrt s} = (7{\rm\ MeV}) \left(R\over 0.01\%\right)
\left(\sqrt s\over 100\rm\ GeV\right) \,,
\end{equation}
where a resolution down to $R=0.003\%$ may be realized at the FMC. In
comparison, a value of $R\sim 1\%$ 
is expected at the Next Linear $e^+e^-$ Collider (NLC). To
study a Higgs resonance one wants to be able to tune the machine energy to
$\sqrt s = m_h$. For this purpose the monochromaticity of the beam energy is
vital.

When the resolution is much larger than the Higgs width, $\sigma_{\sqrt s} \gg
\Gamma_h$, the effective $s$-channel cross section is
\begin{equation}
\bar\sigma_h = \pi \sqrt{2\pi} \
{ {\rm BF} (h\to\mu\mu) \, {\rm BF}(h\to X)\over
 m_h^2} \cdot {\Gamma_h\over \sigma_{\sqrt s} } \,.
\end{equation}
It becomes clear that it would be desirable to get as good
a beam energy resolution as possible because of the factor
$\Gamma_h/\sigma_{\sqrt s}$.
In the other extreme when the resolution is much smaller than the width, 
$\sigma_{\sqrt s} \ll \Gamma_h$, the effective cross section is
\begin{equation}
\bar\sigma_h = 4\pi\, {
{\rm BF}(h\to\mu\mu) \, {\rm BF}(h\to X)\over m_h^2
} \,.
\end{equation}

\begin{figure}[htb]
\centering
\leavevmode
\epsfxsize=3in\epsffile{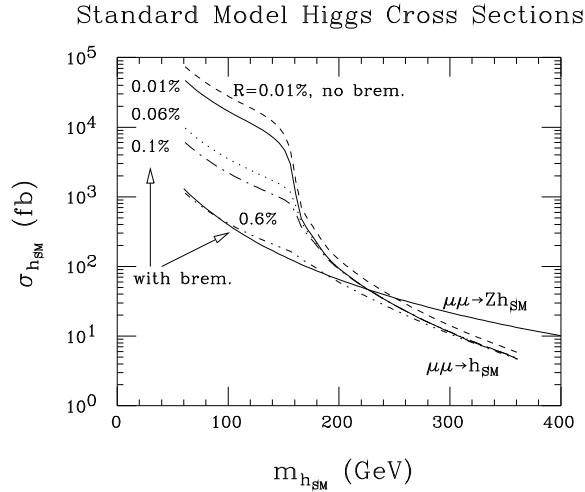}

\caption[]{The $s$-channel cross section for $\mu^+\mu^-\to h$ for
several choices of the beam resolution $R$. Also shown is the $\mu^+\mu^-\to
Zh$ cross section at $\sqrt s = M_Z + \sqrt 2 \ m_h^{}$, from
Ref.~\cite{BBGH}.}
\end{figure}

\begin{figure}[htb]
\centering
\leavevmode
\epsfxsize=3in\epsffile{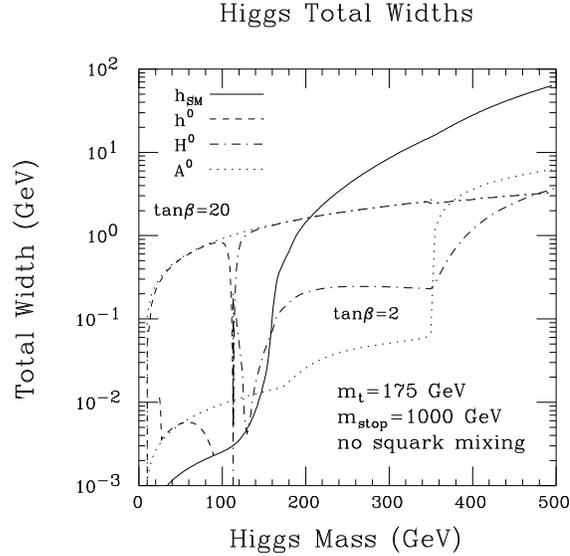}

\caption[]{Total width of the SM and MSSM Higgs bosons with $\tan\beta=2$ and
20, from Ref.~\cite{BBGH}.}
\end{figure}

Figure~1 illustrates the SM Higgs cross section for several choices of the
machine resolution. For $m_h<2M_W$, $\Gamma_h$ is very narrow
and a better beam resolution can significantly improve the signal rate.
On the other hand, for $m_h>2M_W$ the SM Higgs boson becomes increasingly
broad and the effect of $\sigma_{\sqrt s}$ is negligible.
To effectively explore Higgs physics, the resolution requirements 
for the machine thus depend on the Higgs width. 
Figure~2 gives both SM and SUSY Higgs width predictions versus the
Higgs mass. A SM Higgs of mass $m_h\sim100$~\GeVcc \ has a width of a 
few MeV. The
width of the lightest SUSY Higgs may be comparable to that of the SM Higgs (if
$\tan\beta\sim1.8$) or much larger ($\Gamma_h\sim0.5$~GeV for
$\tan\beta\sim 20$). The width parameter characterizes the
fundamental couplings of the Higgs boson to other particles.
Figure~3 shows light Higgs resonance profiles versus
the c.m.\ energy $\sqrt s$. With a resolution $\sigma_{\sqrt s}$ of order
$\Gamma_h$ the Breit-Wigner line shape can be measured and $\Gamma_h$
determined. For the moment we must plan for a resolution $R\lsim0.01\%$ in
order to be sensitive to $\Gamma_h$ of a few MeV.

\begin{figure}[htb]
\centering
\leavevmode
\epsfxsize=3in\epsffile{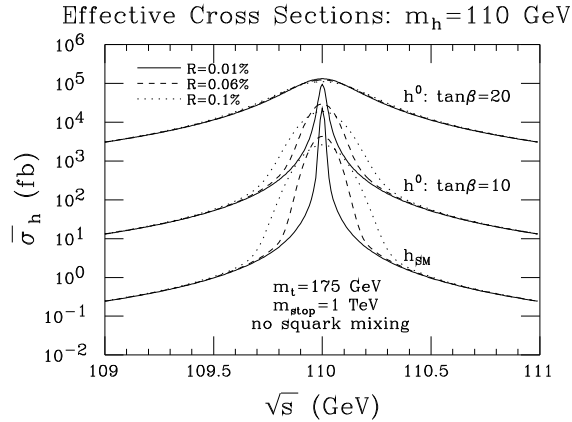}

\caption[]{Effective $s$-channel Higgs cross section $\bar\sigma_h$ obtained by
convoluting the Breit-Wigner resonance formula with a Gaussian distribution for
resolution $R$, from Ref.~\cite{BBGH}.}
\end{figure}

The prospects for observing the SM Higgs are evaluated in Fig.~4. The first two
panels give the signal and background for a resolution $R=0.003\%$. The third
panel gives the luminosity needed for a $5\sigma$ detection in the dominant
$b\bar b$ final state. The luminosity requirements are very reasonable, except
for the $Z$-boson peak region.

\begin{figure}[htb]
\centering
\leavevmode
\epsfxsize=3in\epsffile{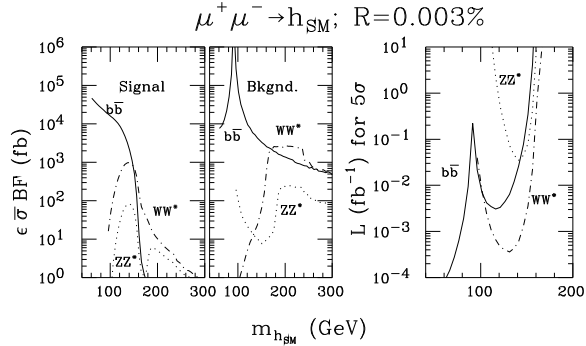}

\caption[]{The SM Higgs cross sections and backgrounds in $b\bar b,\ WW^*$ and
$ZZ^*$. Also shown is the luminosity needed for a 5~standard deviation
detection in the $b\bar b$ decay mode, from Refs.~\cite{Gunion,BBGH}.}
\end{figure}

\looseness=-1
{}It is likely that a SM-like Higgs boson will have been discovered
at the LHC or NLC when the FMC starts its mission. The mass of a light
Higgs boson will have been measured to an accuracy of 
approximately 200~\MeVcc~\cite{CMS}.
From a rough scan for the $s$-channel $h^0$ signal over this 200 MeV
range, the mass can be determined
to an accuracy $\Delta m_h \sim\sigma_{\sqrt s}$. If $S/\sqrt B\gsim3$ is
required for detection or rejection of a Higgs signal and a resolution
$R\sim0.003\%$ ($\sigma_{\sqrt s} \sim 2$~MeV) is employed, then the necessary
luminosity per scan point is 0.0015~fb$^{-1}$ for $m_h \lsim 2M_W$ and $m_h$
not near $M_Z$. As an example, suppose that the LHC has measured 
$m_h = 110.0 \pm 0.1$~\GeVcc. 
The number of scan points to cover a 200~\MeVcc \ region in
$\sqrt s$ at the FMC is $\sim \rm200~MeV/2~MeV
= 100$, and a total luminosity of $100\times (0.0015\rm~fb^{-1}/point) =
0.15~fb^{-1}$ is needed to discover the Higgs and reach an accuracy on its mass
of
\begin{equation}
\Delta m_h \simeq \sigma_{\sqrt s} \sim 2\rm\ \MeVcc \,.
\end{equation}

Once $m_h$ is determined to an accuracy $\Delta m_h \sim {\cal O}
\left(\sigma_{\sqrt s}\right)$ a three point fine scan can be made with one
setting at the apparent peak and two settings on the wings at $\pm\sigma_{\sqrt
s}$ from the peak. The ratios of $\sigma({\rm wing}^i)/\sigma({\rm peak}^i)$
determine $m_h$ and $\Gamma_h$. With a good energy resolution $R=0.003\%$
and an integrated luminosity $L_{\rm total} = 0.4\rm~fb^{-1}$,
for $m_{h}=110$~\GeVcc \ and $\Gamma_{h}=3$~MeV, accuracies of 
\begin{equation}
\Delta \Gamma_h/\Gamma_h = 16\% \, ,\quad {\rm and}\, \quad
\Delta m_h \sim 0.1\, {\rm \MeVcc}
\end{equation}
could be achieved. At the same time, branching ratios
for the dominant decay channels can be also measured
to a good precision, {\it i.e.}, $3\%$ and $15\%$
for $\sigma \cdot {\rm BF}(b\bar b)$ and 
$\sigma \cdot {\rm BF}(W^+W^-)$, respectively.
The high precision reached would have significant impact
on the electroweak physics within and beyond the SM. 
For instance, with a measurement of 
an accuracy about $15\%$ on the ratio
${\rm BF}(W^+W^-)/{\rm BF}(b\bar b)$, 
one should be able to infer $A^0$ effects up to 
$M_{A^0}\gsim 400$~\GeVcc\  by comparing the predictions 
from the MSSM and the SM \cite{Gunion,BBGH}.
However, to reach the necessary precision within 
a sensible time scale, a machine luminosity of
$10^{32}$ cm$^{-2}$s$^{-1}$ (or 1 fb$^{-1}$/yr) 
and a good energy resolution $R=0.003\%$ are 
highly desirable.

The heavier neutral MSSM Higgs bosons are also observable in the $s$-channel.
Figure~5 give the cross sections and significance of the ${\cal CP}$-odd 
state $A^0$
versus the $A^0$ mass, assuming $R=0.1\%$ and $L=0.01\rm~fb^{-1}$. Discovery
and study of the $A^0$ is possible at all $m_A$ if $\tan\beta>2$ and at
$m_A<2m_t$ if $\tan\beta\lsim2$.

\begin{figure}[htb]
\centering
\leavevmode
\epsfxsize=3.1in\epsffile{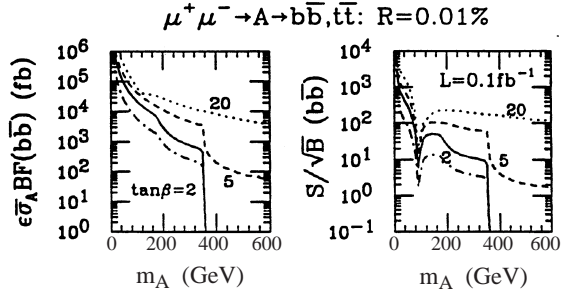}

\caption[]{Cross sections and significance for detection of the $A$ Higgs boson
with an efficiency $\epsilon=0.5$ and a luminosity $L=0.1\rm~fb^{-1}$, from
Ref.~\cite{BBGH}.}
\end{figure}

The possibility that $A^0$ and $H^0$ may be nearly mass degenerate is of
particular interest for $s$-channel Higgs studies. In the large $m_A$ limit,
typical of many supergravity models, the masses of $A^0$, $H^0$ and $H^\pm$ are
similar and $h^0$ is similar to $h_{\rm SM}$ in its properties. In this
situation the $A^0$ and $H^0$ contributions can be separated by an $s$-channel
scan; see Fig.~6.


\begin{figure}[htb]
\centering
\leavevmode
\epsfxsize=3.5in\epsffile{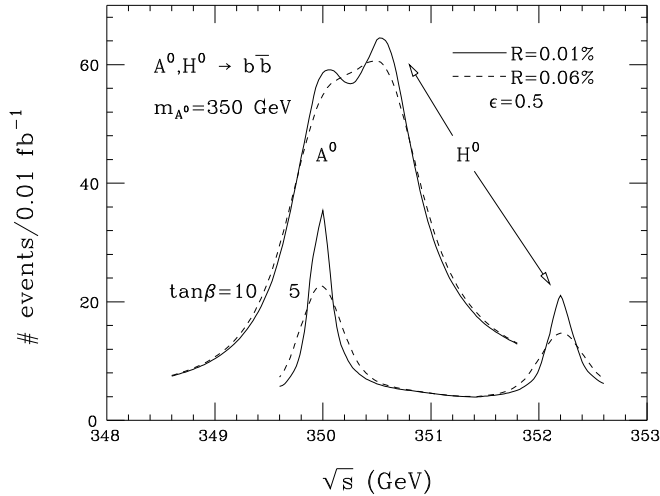}

\caption[]{Separation of $A^0$ and $H^0$ Higgs signals for two values of
$\tan\beta$, from Ref.~\cite{BBGH}.}
\end{figure}

It was reported during the workshop \cite{KMP}
that beam polarization is potentially useful for Higgs
resonance studies, but only if the accompanying luminosity 
reduction is not significant. Large forward-backward asymmetries 
can also be used to enhance the Higgs ``discovery'' signal or 
improve precision measurements, particularly for the $\tau\bar{\tau}$ 
final state. 

If the muon collider is running at energies above the narrow
Higgs resonance, it may still be possible to pick up a 
signal sample through the photon radiation process
$\mu^+\mu^- \to \gamma h$ with a cross section of the order of 
0.1~fb. The authors in \cite{Bowser-Chao}
studied this process and pointed out that the 1-loop contribution
is comparable in size to the tree level result, and is
especially sensitive to the Higgs coupling to the top quark
and to anomalous Higgs couplings to $W^+W^-$ and $ZZ$. 
It is also discussed~\cite{Keung}
that ${\cal CP}$-odd kinematic variables may be 
constructed for processes like $h^0 \to W^+W^-,ZZ$ and $t\bar t$
so that one may be able to probe the ${\cal CP}$ properties of the
Higgs boson couplings at muon colliders.

During the workshop, many other aspects of Higgs physics
were discussed. They include Higgs boson searches at
LEP \cite{Greening,Maria}; Higgs physics at the Tevatron
within the SM \cite{John} and within the MSSM \cite{Steve},  at
the LHC \cite{Kao,Dave}, and at the NLC \cite{Han0}.

\section*{$Z$ Pole Electroweak Physics}

The success of the $\SM$ has arguably been most beautifully demonstrated
by the agreement of the very precise $Z$ pole measurements at the 
$\ee$ colliders and the direct measurements. It has
been unprecedented that an anticipated quark was discovered with
a mass exactly within the range predicted from loop corrections within 
a theoretical framework. 
This is a remarkable feat for experimentalists and theorists alike 
and attests to the enormous success of the $\SM$. 
Even though many measurements are now being carried out with excruciating 
precision, the $\SM$ shows no signs of
giving up its claim of being the description of the fundamental
interactions as we know them. 
Despite these enormous successes there are some discrepancies in the data 
which, given that the LEP $Z$ pole era is over, will most likely stay with us 
for a long time. 
The most significant discrepancy from lepton colliders is the 
measurement of $\swsqeffl$, defined as 
\begin{displaymath}
    \swsqeffl   \,\equiv\, {1\over 4} \, 
                           \left( 1 - {{g_V^f}^2 \over {g_A^f}^2} 
                           \right) \ \ , 
\end{displaymath} 
where $g_{V(A)}^f$ is the (axial-)vector coupling of the $Z$ boson to 
fermion $f$. Currently SLD measures 
$\swsqeffl = 0.23055 \pm 0.00041$~\cite{willocq}, 
derived from the left-right asymmetry 
of the total cross section and the leptonic forward-backward asymmetries, 
compared to the LEP average of 
$\swsqeffl = 0.23196 \pm 0.00028$, determined from the $Z$ partial
decay widths and the forward-backward asymmetries~\cite{ewwg}. 
The discrepancy between the measurements has a significance of  
2.8$\sigma$. 
This discrepancy is rather significant. 
Considering only the $\ALR$ measurement, 
combined with the direct determination of the $W$ and top quark masses, 
the measurement implies a 95\% CL upper bound on the Higgs mass of 
77~GeV/c$^2$, while the direct searches at LEP yield a 
95\% CL lower limit on 
the Higgs mass also of 77~GeV/c$^2$~\cite{chanowitz}. 

The FMC with its anticipated luminosity could contribute significantly 
to the physics in this sector~\cite{blondel}. 
Within a one year running period, due to a relaxed requirement
on the beam energy resolution $R$ for the broad $Z$-pole, a sample 
of $10^8$ $Z$ events could be recorded with both beams naturally polarized. 
To gauge the possible impact of such a large data sample it is instructive to 
look at the sensitivity of electroweak observables to $\SM$ 
parameters. Table~\ref{tab:ewk_sens} lists electroweak observable 
${\cal O}$ 
together with its current measurement and sensitivity, 
$\Delta {\cal O}$,  
to the top quark mass, $m_t = 175.6 \pm 5.5~\GeVcc$,
the fine structure constant, 
$\alpha^{-1}(m_Z^2) = 128.896 \pm 0.090$, 
the strong coupling constant,  
$\alpha_s(m_Z^2) = 0.121 \pm 0.003$, and 
the Higgs boson mass, $60 < m_h < 1000~\GeVcc$~\cite{zfitter}.
The observables chosen are 
$\Rl$, the ratio of the hadronic over leptonic partial decay width of the 
$Z$, $\Rl = \Gll / \Ghad$;  
$\Gll$, the $Z$ partial decay width into leptons;
$\swsqeffl$; 
$\Rb$, the fraction of hadronic $Z$ decays coming from $b$ quarks, 
$\Rb = \Gbb / \Ghad$; and the mass of the $W$ boson, $m_W$. 
The choice of these observables is given by their experimental uncertainty 
compared to their sensitivity to the various parameters in the model.

\begin{table}[ht]
\begin{center}
\begin{tabular}{lr@{$\ \pm\ $}ldddd} 
\noalign{\vspace{-8.0pt}}
\tableline
  Observable (${\cal O}$) \qquad 
	     & \multicolumn{2}{c}{Average Value \quad } 
	     & $\Delta {\cal O}(\delta M_t)$   
	     & $\Delta {\cal O}(\delta\alpha)$      
	     & $\Delta {\cal O}(\delta\alpha_s)$    
	     & $\Delta {\cal O}(\delta m_h )$       \\[4pt]
\hline 
\noalign{\vspace{3.0pt}}
  $\Rl \cdot 10^3$  	   & 20755  &   27 & 1.8  & 4.0 & 21   & 28 	\\
  $\Gll$ (\MeV)	    	   & 83.91  & 0.10 & 0.06 & --- & 0.02 & 0.25 	\\
  $\swsqeffl \cdot 10^4 $  & 2315.2 & 2.3  & 2.0  & 2.3 & 0.05 & 15.4 	\\ 
  $\Rb \cdot 10^4 $        & 2170   &  9   & 2.0  & 0.2 & 0.05 & 0.4 	\\ 
  $m_W$ (\MeV/c$^2$) \quad & 80430  & 80   & 37   & 14  & 1    & 200 	\\
\noalign{\vspace{3.0pt}}
\hline 
\end{tabular}
\caption[]{Results for various electroweak observables 
           and their sensitivity to the top quark mass, $\alpha$, 
	   $\alpha_s$ and the Higgs boson mass. } 
\label{tab:ewk_sens}
\end{center}
\end{table}

From the table one should observe that $\Gll$ is insensitive to 
$\alpha$, whereas $\Rl$ is very sensitive to $\alpha_s$. 
It is also clear that the constraint on the Higgs mass is 
dominated by the measurement of $\swsqeffl$. 
The measurement of $M_W$ will become as significant as the current 
measurement of $\swsqeffl$ when the experimental accuracy reaches 
a level of 30~\MeVcc. If at that point, however, the top mass 
uncertainty has not been reduced, the constraint from the $M_W$ measurement 
would partially be spoiled by the top mass uncertainty.

Within the framework of the $\SM$ the value of $\alpha_s(m_Z^2)$ derived 
from an analysis of electroweak precision data depends essentially on 
$\Rl$, $\GZ$ and $\shad$, with $\shad$ the peak hadronic $Z$ pole cross
section. 
Since $\Rl$ is very sensitive to $\alpha_s$,  
the strong coupling parameter can be determined from the parameter $\Rl$ 
alone. For $m_Z = 91.1867$~\GeVcc, and imposing 
$m_t= 175.6 \pm 5.5$~\GeVcc \ as a constraint, a value of
$\alpha_s(m_Z^2) = 0.124 \pm 0.004 \pm 0.002$ is obtained, where the second 
uncertainty accounts for the change in the result when varying $m_h$ in the 
range $60 < m_h < 1000$~\GeVcc~\cite{ewwg}. 
The experimental uncertainty is dominated by the limited statistics of the 
leptonic $Z$ decays. With an improvement in statistics of a factor of 
10 over the current LEP statistics, an uncertainty on $\alpha_s$ of 0.001 
could be obtained~\cite{blondel}. 


The quantity $\Gll$ is of particular interest since it is independent 
of $\alpha$ and is only mildly dependent on $\alpha_s$. As such, it 
is a sensitive indicator of possible new physics beyond the $\SM$. 
A precision measurement 
of $\Gll$, however, requires a very accurate absolute luminosity calibration. 
$\Gll$ can be determined with better precision indirectly from 
\begin{displaymath}
    \GZ \,=\, \Gll \, ( \, 3 \,+\, N_\nu \, {\Gnu \over \Gll} \,+\, \Rl \, ) 
\end{displaymath} 
using the measurement of $\GZ$, the $\SM$ prediction for 
${\Gnu \over \Gll}$ and the measurement of $\Rl$. 
The uncertainty on $\Rl$, as seen above, will be very small and 
independent of luminosity. The uncertainty is driven by the uncertainty 
on the $Z$ width, which depends on the luminosity through the point to 
point errors in the energy scan and on 
the energy calibration. At the FMC, where a continuous energy calibration 
should be feasible, the latter can be considerably reduced and an 
accuracy of ${\delta\Gll \over \Gll } = 0.0003$ should be possible, 
compared to the current measurement of $\Gll = 83.91 \pm 0.10$.

Currently the most powerful way to determine $\swsqeffl$ is to measure the 
left-right asymmetry, $\ALR$, defined as
\begin{eqnarray}
    \ALR \,&=&\, -{ 1 \over P} \, 
		    { \sigma_{\rm L} \,-\, \sigma_{\rm R} \over 
		      \sigma_{\rm L} \,+\, \sigma_{\rm R} }
\label{eq:alr}
\end{eqnarray}
where 
$\sigma_{R(L)}$ is the total production cross section for a
right (left) handed polarized electron beam with average 
polarization $P$.  
At the $Z$ pole, ignoring photonic corrections, $\ALR = \cAe $, 
where the asymmetry of couplings, $\cAf$, is given by 
\begin{displaymath}
    \cAf    \,\equiv\, { {g_L^f}^2 \,-\, {g_R^f}^2  \over 
                         {g_L^f}^2 \,+\, {g_R^f}^2  } 
            \,=\,      {2\, g_V^f \, g_A^f \over 
                           {g_V^f}^2 \,+\, {g_A^f}^2 } \ \ .
\end{displaymath} 
The measurement of $\ALR$ at SLD, using a polarized electron beam with 
an average polarization during the last run of 
$ {\cal P}_e = (76.5 \pm 0.8)\%$, yields a measurement of 
$\swsqeffl = 0.23055 \pm 0.00041$~\cite{willocq}, 
of comparable precision to the combined LEP result, derived from the 
$Z$-pole and $\AFB$ measurements. 
The power lies in the availability of polarized beams. 
The sensitivities of the two measurements are related as 
${\partial\ALR / \partial\swsqeffl} = {1\over 4} 
 {\partial\AFB / \partial\swsqeffl} $. 
Thus, compared to an $\ALR$ measurement with fully polarized beams, a
sixteen-fold larger data sample is required to achieve a similar
accuracy in $\swsqeffl$ from $\AFB$.

When both beams are polarized, the natural situation for a muon collider, 
equation~\ref{eq:alr} generalizes to 
\begin{eqnarray}
    \ALR \,&=&\,  { 1 \over {\cal P}} \, 
		  { \sigma_{\rm L} \,-\, \sigma_{\rm R} \over 
		    \sigma_{\rm L} \,+\, \sigma_{\rm R} } 
                    \qquad {\rm with} 				\\
    {\cal P} &=&  {P^+ \,-\, P^-  \over 
		   1 \,-\, P^+ \,P^-  }
\label{eq:genalr}
\end{eqnarray}
where $P^{+(-)}$ refers to the average longitudinal polarization of the 
positively (negatively) charged incident fermion beam.  
The total cross section is given by 
\begin{eqnarray}
    \sigma \,&=&\, \sigma_0 \, 
		   \{ \,
		      ( 1 \,-\, P^+P^-) \,+\, (P^+ \,-\, P^-) \, \ALR
		   \}  \ .
\end{eqnarray}
From equation~\ref{eq:genalr} it can be seen that the quantity that 
controls the measurement of 
$\ALR$ (=$\cAe$), and thus the precision of the $\swsqeffl$ measurement, is 
${\cal P}$, the effective polarization of the $\mu^+\mu^-$-system, which 
enhances the sensitivity of the measurement. For a 50\% polarization 
of both beams, $P^\pm = \pm 50\%$, ${\cal P} = 80\%$. 
On top of that the cross section also increases by about 30\%.
Given the process under study a signal to background enhancement can be 
obtained by optimizing the polarization~\cite{KMP}. The most clear 
example is the production of the scalar Higgs boson, where the same sign 
spin is the favored production mode with a strong suppression of the 
background.

\begin{figure}[t]
	\epsfxsize = 8.0cm
	\centerline{\epsffile{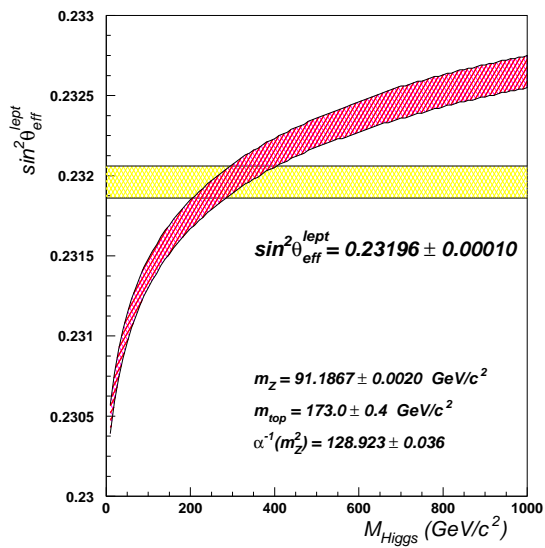}}
\vspace{10pt}
\caption[]{Dependence of $\swsqeffl$ on $m_h$ and the constraint on $m_h$ 
	   a hypothetical measurement of $\swsqeffl$ at the $10^{-4}$ level 
	   would yield, assuming an uncertainty on $m_t$ of 2.0~\GeVcc . }
\label{fig:seff_mh}
\end{figure}

In the current design of the FMC a big loss in luminosity is incurred 
for increased polarization~\cite{fmc}. For a polarization of 50\% 
there is a loss in luminosity of about a factor of 4 per beam. 
The statistical precision with which $\ALR$ can be measured is 
\begin{eqnarray}
    \delta \ALR \,&=&\, {1 \over {\cal P} } \, 
			{1 \over {\sqrt N}} \ .
\end{eqnarray}
The relevant quantity therefore to collect the data to measure $\ALR$ with 
a certain precision is $L_P \cdot {\cal P}^2$, where 
$L_P$ is the loss in luminosity for beam polarization $P$~\cite{blondel}.  
Given the relation between the loss in luminosity and the beam polarization 
in the current design of the FMC, this quantity reaches a maximum for a loss 
in luminosity of about 0.7 corresponding to a beam polarization of about
30\%. 
The advantage of fully polarized beams does not outweigh the loss in 
luminosity incurred in the current design. 
The uncertainties on $\ALR$ and $\swsqeffl$ are 
related through $ \delta \ALR = 7.9 \, \delta\swsqeffl$. A data sample 
of $10^7$ $Z$ events with ${\cal P} = 0.5$, assuming equal statistical and 
systematic uncertainties, should make a measurement of 
$\delta\swsqeffl < 10^{-4}$ easily feasible. The power of a measurement 
of $\swsqeffl$ to such a precision can readily be seen from 
Fig.~\ref{fig:seff_mh} which shows the dependence of $\swsqeffl$ on $m_h$ 
taking as input the current measured value for $m_Z$ and assuming 
$m_t = 173.0 \pm 0.4$~\GeVcc and 
$\alpha^{-1}(m_Z^2) = 128.923 \pm 0.036$~\cite{davier}. 
Taking the current LEP central value for $\swsqeffl$ a precision of 
$10^{-4}$  would constrain the Higgs mass to the range 
$170 < m_h < 370$~\GeVcc.  It should be noted that the uncertainty on the 
$\SM$ prediction is dominated by the uncertainty on $\alpha$. 
An error on $m_t$ of 2~\GeVcc \ is equivalent to an uncertainty on $\alpha$ 
of 0.04. With further improvements to the 
hadronic contribution of $\alpha$ to be anticipated, a precise measurement 
of $\swsqeffl$ combined with a direct measurement of the Higgs mass, 
provides a very stringent consistency check of the $\SM$.

It should be noted that relative total cross section measurements with 
different spin configurations also gives a measure of the polarization. 
This measures directly the polarization of the interacting muons and no 
corrections for beam transport to a polarimeter and sampled luminous region 
need to be applied. 

Many more $Z$ pole quantities can be studied at the FMC, notably in the 
b sector. This, however, has at present not been investigated and should 
be explored in future studies.

\section*{$W$ Mass}

\begin{figure}[t]
	\epsfxsize = 8.0cm
	\centerline{\epsffile{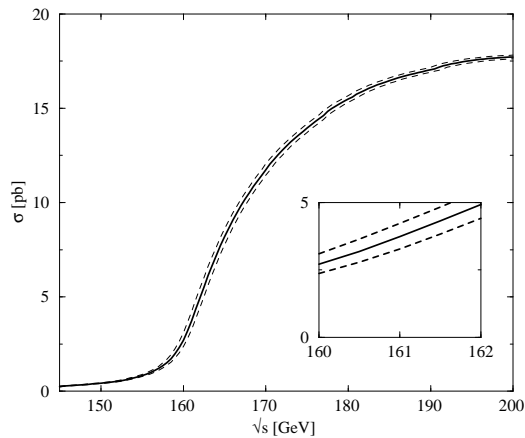}}
\vspace{10pt}
\caption[]{The cross section for $\mu^+\mu^- \rightarrow W^+W^-$ for 
	   $m_W = 80.3 \ (\pm 0.2)$~\GeVcc \ indicated by the solid (dashed) 
	   lines.  The inset shows the cross section in the region of maximum 
	   statistical sensitivity. }
\label{fig:ww_thresh}
\end{figure}

Until very recently the mass of the $W$ boson could only be measured
directly in $\pbarp$ collisions. 
Precise measurements of the $W$ mass have recently been obtained at 
LEP2~\cite{ewwg} 
using the enhanced statistical power of the rapidly varying total
$W^+W^-$ cross section at threshold, and the 
Breit-Wigner peaking behavior of
the invariant mass distribution of the $W$ decay products. 
By measuring the $WW$ threshold cross section at 
$\sqrt{s}=161$~\GeV, the four LEP experiments have
obtained a combined $W$ mass value of 
$m_W = 80.40^{+0.22}_{-0.21} \pm 0.03$~\GeVcc. 
The second error is due to the LEP energy calibration. 
The direct reconstruction of the $W$ mass from the decay products gives 
$m_W = 80.37 \pm 0.18 \pm 0.05 \pm 0.03$~\GeVcc. The second uncertainty 
is due to effects of color reconnection and the last error is due to the 
LEP energy calibration. The achievable precision on $m_W$ from LEP 
for an integrated luminosity of 500~pb$^{-1}$ per experiment is 
estimated to be $\delta m_W = 35$~\MeVcc~\cite{lep2_yellow}. 
For an integrated luminosity of 10~fb$^{-1}$ the Tevatron might be 
able to constrain $m_W$ to about 20~\MeVcc~\cite{snowmass}. 
The FMC is particularly well suited to a threshold measurement because 
of the very narrow beam spread, the ability to determine the beam energy 
during a physics run, and the reduced initial state radiation (ISR). 
Due to the large backgrounds, however, systematic errors arising 
from uncertainties in both the background level as well as the detection 
efficiencies will limit the ultimate precision~\cite{han,berger}. 
The dominant physics background is mainly due to $Z\gamma$   
production, which is almost independent of energy.  
When detection efficiencies and backgrounds, including the beam induced 
backgrounds, are largely energy independent, 
the best precision is obtained through a ratio of cross section measurements
at an energy well below the $WW$ threshold and at the threshold. 
The optimal threshold energy is also for the FMC\cite{lep2_yellow} 
\begin{equation}
	{\sqrt s} \,\approx\, 2 \, m_W \,+\, 0.5~\GeV	\ .
\end{equation}
Figure~\ref{fig:ww_thresh} shows the cross section for 
$\mu^+\mu^- \rightarrow W^+W^-$ in the threshold region for three different 
$W$ mass values. ISR effects have been included. 
With an integrated luminosity of ${\cal L} = 100$~pb$^{-1}$ a precision 
of $\delta m_W = 6~\MeVcc$ from the threshold measurement can be obtained 
when combining the three decay channels $\qqbar\qqbar$, $\qqbar\ell\nu$ and 
$\ell\nu\ell\nu$. 
A 10~\MeVcc \ uncertainty on $m_W$ is equivalent to about a 0.5\% uncertainty 
on the measured cross section.

\section*{Triple and Quartic Gauge Boson Couplings}

The non-Abelian $SU(2)\times U(1)$ gauge symmetry of the $\SM$ implies 
that the gauge bosons self-interact. These self-interactions give rise 
to very subtle interference effects in the $\SM$. In fact, in the $\SM$  
the couplings are uniquely determined by the gauge symmetry in
order to preserve unitarity. 
An accurate measurement of the gauge boson self-interactions would 
constitute a stringent test of the gauge sector of the $\SM$ and any 
observed deviation of the couplings from their $\SM$ value would indicate
new physics. 

The formalism of effective Lagrangians is used to describe gauge boson 
interactions beyond the $\SM$. 
The most general effective electroweak Lagrangian 
contains $2\times7$ free parameters~\cite{Hagiwara}:
$	g_1^V, 	\kappa_V, 	\lambda_V,  
  	g_4^V, 	g_5^V,  
     	\widetilde{\kappa}_V, 	\widetilde{\lambda}_V , 
$
with $ V=\gamma,Z $. 
The parameter  
$       g_5^V, $ violates ${\cal C}$ and ${\cal P}$ but conserves ${\cal CP}$;
$ 	g_4^V, 	
        \widetilde{\kappa}_V$ and $ \widetilde{\lambda}_V$ 
violate ${\cal CP}$. 
In the $\SM$ 
$g_1^V=1, \kappa_V=1, $  and all other parameters vanish. 
For these two parameters one therefore introduces 
deviations from the $\SM$ values, 
$\Delta\kappa_V = \kappa_V - 1$ and 
$\Delta g_1^V   = g_1^V - 1 $.

Gauge boson self-interactions can be studied through di-boson production. 
The cross sections for di-boson production are generally rather small and 
a study of the full fourteen-dimensional parameter space is impossible. 
In general, two approaches are followed to reduce the parameter space. 
The \pbarp experiments generally set all parameters but two to their 
$\SM$ values and concentrate on 
$\Delta\kappa_V$, $\lambda_V$ because they have a direct physical 
connection through the magnetic dipole and electric quadrupole moment of 
the $W$ boson,
$    \mu_W = (e/2m_W)(1+\kappa_\gamma+\lambda_\gamma) $ and 
$    Q_W^e = (-e/m_W^2)(\kappa_\gamma-\lambda_\gamma) $~\cite{Kim}. 

The second approach, followed mainly by the LEP experiments, constructs
an effective Lagrangian with operators of higher dimension.
By imposing some restriction, like retaining only the lowest dimension 
operators, respecting 
${\cal C}, {\cal P}$ and ${\cal CP}$ invariance and requiring the 
Lagrangian to be invariant under 
$SU(2)\times U(1)$ and adding a Higgs doublet, the number of free 
parameters is reduced to just three~\cite{anomalous_lep}.
With further, rather ad hoc, requirements the parameter space can be 
reduced to just two free parameters, with definite
relations between the different parameters~\cite{hisz}. 

If in the processes of di-boson production 
the couplings deviate even modestly from their 
$\SM$ values, the gauge cancellations are destroyed and a large increase 
of the cross section is observed. Moreover, the differential distributions 
will be modified.
A $WWV$ interaction Lagrangian with constant anomalous couplings
would thus violate unitarity at high energies and therefore
the coupling parameters are modified to include
form factors~\cite{Baur}, that is, 
$\Delta\kappa (\hat{s}) = \Delta\kappa/(1+\hat{s}/\Lambda^2)^2 $ and 
$      \lambda(\hat{s}) =      \lambda/(1+\hat{s}/\Lambda^{2})^{2}$, 
where $\hat{s}$ is the square of the center of mass energy of the
subprocess. 
$\Lambda$ is a uni\-tarity preserving form factor scale 
and indicates the scale at which new physics would manifest itself. 

Currently the strongest limits come from the D0 experiment.
From a combined fit to the results from the  
$WW$, $WW/WZ$ and $W\gamma$ analyses based on the full  Run~I data, 
the limits obtained at 95\% CL are ($\Lambda = 1.5$~TeV):  
\begin{center}
\begin{tabular}{cc}
    $-0.33 < \Delta\kappa < 0.45$    & $(\lambda = 0)$         \\
    $-0.2 < \lambda < 0.2$           & $(\Delta\kappa = 0)$,
\end{tabular}
\end{center}
where it was assumed that the $WWZ$ couplings and the $WW\gamma$
couplings were equal~\cite{win97}. 
A relatively small improvement in the limits is anticipated for the 
combined Tevatron data. 
With 500~$pb^{-1}$ at $\sqrt s = 190$~\GeV\ limits of the order of 
0.05 to 0.1 on anomalous couplings are expected from LEP~\cite{lep2_yellow}. 
At the NLC, for a center of mass energy of ${\sqrt s} = 500$~\GeV\
limits of 
$|\Delta\kappa_\gamma | < 2.4 \cdot 10^{-3}$ and 
$|\lambda_\gamma | < 1.8 \cdot 10^{-3}$ are expected~\cite{nlc}.
The clear advantage of the FMC is the reduced initial state radiation (ISR), 
which will facilitate the event reconstruction and reduce the systematic 
uncertainties. No significant improvement in the limits over the constraints 
from an NLC are expected~\cite{baur1}.

On the other hand, it was reported~\cite{HE}
at the workshop that both the NLC and a high energy muon collider
with $\sqrt s = 0.5-1.5$~TeV may have
significant sensitivity to perform direct tests 
of the quartic gauge boson couplings
to a precision of theoretical interests.
With an integrated luminosity of 200~fb$^{-1}$ limits on the anomalous 
quartic couplings $\ell_4$, $\ell_5$, $\ell_6$, $\ell_7$ and $\ell_{10}$ 
of ${\cal O}(10^{-1})$ will be possible. Also here, beam polarization 
significantly improves the sensitivity to anomalous couplings.

\section*{Conclusions}

There is a rich physics program available with a muon collider. 
Among them, $s$-channel Higgs boson production may prove
to be the ``Crown jewel'': precision measurements on the Higgs
mass, width, decay branching fractions and couplings to other
particles may provide invaluable information on physics within
and beyond the SM. However, extremely good beam energy resolution
($R=0.003\%$) and a high luminosity ($10^{32}$cm$^{-2}$s$^{-1}$)
are highly desirable to fulfill the physics goal.
$Z$ pole physics will yield significant results if a data sample 
of $10^8$ $Z$ events
(corresponding to an integrated luminosity of about 2 fb$^{-1}$)
can be recorded with polarized beams.
Spin manipulation is extremely powerful if it is built-in 
in the machine design from the start.

\section*{Acknowledgments}
 We would like to thank the Workshop organizers for the invitation to
the working group, and would like to thank the participants of the 
working group \cite{part} for their contributions. 

\end{document}